\documentclass[%
superscriptaddress,
preprint,
showpacs,
 amsmath,amssymb,
 aps,
pre
]{revtex4-1}

\usepackage{graphicx}
\usepackage{dcolumn}
\usepackage{bm}
\usepackage{hyperref}

\allowdisplaybreaks 

\begin{document}


\title{No-boarding buses: Agents allowed to cooperate or defect}

\author{Vee-Liem Saw}
\email{Vee-Liem@ntu.edu.sg}
\affiliation{Division of Physics and Applied Physics, School of Physical and Mathematical Sciences, 21 Nanyang Link, Nanyang Technological University, Singapore 637371}
\affiliation{Data Science and Artificial Intelligence Research Centre, Block N4 \#02a-32, Nanyang Avenue, Nanyang Technological University, Singapore 639798}
\author{Lock Yue Chew}
\email{lockyue@ntu.edu.sg}
\affiliation{Division of Physics and Applied Physics, School of Physical and Mathematical Sciences, 21 Nanyang Link, Nanyang Technological University, Singapore 637371}
\affiliation{Data Science and Artificial Intelligence Research Centre, Block N4 \#02a-32, Nanyang Avenue, Nanyang Technological University, Singapore 639798}
\affiliation{Complexity Institute, 61 Nanyang Drive, Nanyang Technological University, Singapore 637335}
%

\date{\today}

\begin{abstract}
We study a bus system with a no-boarding policy, where a ``slow'' bus may disallow passengers from boarding if it meets some criteria. When the no-boarding policy is activated, people waiting to board at the bus stop are given the choices of \emph{cooperating} or \emph{defecting}. The people's heterogeneous behaviours are modelled by inductive reasoning and bounded rationality, inspired by the El Farol problem and the minority game. In defecting the no-boarding policy, instead of the minority group being the winning group, we investigate several scenarios where defectors win if the number of defectors does not exceed the maximum number of allowed defectors but lose otherwise. Contrary to the classical minority game which has $N$ agents repeatedly playing amongst themselves, many real-world situations like boarding a bus involves only a subset of agents who ``play each round'', with \emph{different subsets playing at different rounds}. We find for such realistic situations, there is no phase transition with no herding behaviour when the usual control paramater $2^m/N$ is small. The absence of the herding behaviour assures feasible and sustainable implementation of the no-boarding policy with allowance for defections, without leading to bus bunching.
\end{abstract}

\maketitle


\section{Introduction}

Bus transit systems play a vital role in moving people efficiently. Left on its own, however, buses tend to bunch and form clusters of buses moving together. The formation of such clusters can be intuitively understood as follows: Suppose buses are initially distributed evenly. Due to stochasticity in the number of people waiting at a bus stop as well as traffic conditions, a bus may happen to be slightly delayed at a bus stop. Once it leaves the bus stop, the bus that is trailing it experiences a slightly shorter headway such that it needs to pick up slightly less people than the original bus. This speeds up the trailing bus further, allowing it to catch up even more with fewer and fewer people at the subsequent bus stops for it to pick up due to the diminishing headway. Therefore, these two buses end up bunching together. Bunched buses reduce the efficacy of the system, because if a person misses a group of bunched buses, he essentially misses not one but \emph{multiple} buses and have to wait longer for the next bus(es) to arrive.

This problem has long been identified \cite{Newell64,Chapman78,Powell83,Gers09,Bell10,Vee2019}, with many studies conducted to propose possible rectifications. Some suggested methods include holding back buses so that they follow a prescribed schedule or to counteract the diverging headways between buses \cite{Abk84,Ros98,Eber01,Hick01,Fu02,Bin06,Daganzo09,Cor10,Cats11,Gers11,Bart12,Moreira16,Wang18}, stop-skipping to correct for the headways \cite{Li91,Eber95,Fu03,Sun05,Cor10,Liu13}, deadheading (i.e. having an empty bus move directly to a designated bus stop) \cite{Furth85,Furth85b,Eber95,Eber98,Liu13}, carefully engineering the bus routes and locations of the bus stops \cite{Tirachini14}, using buses with wide doors to speed up boarding/alighting \cite{WideDoor,Steward14,Geneidy17}, as well as a no-boarding policy where a ``slow'' bus only allows people to alight but disallows boarding \cite{Vee2019b}.

\subsection{No-boarding buses}

In the no-boarding policy, the ``slow'' bus gets to speed up by saving time from otherwise getting stuck at the bus stop to pick up people, and allows the ``fast'' bus behind it (which would soon bunch into it, if no interference is carried out) to pick up these people instead --- effectively slowing it down. Ref.\ \cite{Vee2019b} has worked out analytically, backed up by extensive simulations based on a real bus system, that the bus system as a whole would experience significant improvement in terms of reducing the overall average waiting time of people at the bus stop for a bus to arrive. The global improvement comes with minor local cost, however, as those denied boarding would doubtlessly have their own waiting times slightly extended. Nevertheless, the overall global gain far outweighs those minor local costs.

\begin{figure}
\centering
\includegraphics[width=10cm]{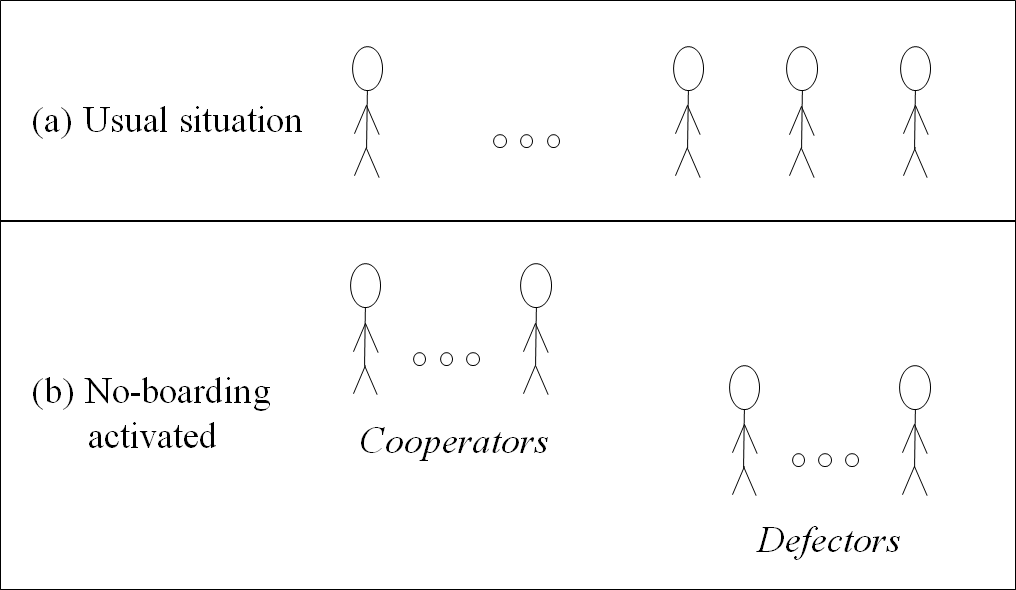}
\caption{(a) The usual situation where a queue of people at a bus stop boards a bus. (b) The bus activates ``no-boarding'': Each person in the queue is given the choices to cooperate or defect the no-boarding rule. The cooperators remain at the bus stop and wait for the next bus, whilst the defectors proceed as usual. If cooperators win, they receive a rebate with the defectors paying a fine. On the other hand if defectors win, they get away without paying a fine with the cooperators not receiving any rebate.}
\label{fig1}
\end{figure}

The purpose of this paper is to investigate the scenario where unlike in Ref.\ \cite{Vee2019b} with no-boarding mandatorily enforced by the bus system when a bus is deemed as ``too slow'', here the people waiting to board at the bus stop are given the options of whether to \emph{cooperate} or \emph{defect} the no-boarding policy (see Fig. \ref{fig1}, and also Section 3). Such a social situation is of immense interest to policymakers and bus operators because sometimes some people necessarily require service urgently and are willing to pay a premium for it --- of course, provided that such an option is available in the first place. On top of that, certain weather conditions like thunderstorm, snow (in countries with winter), blazing heat, amongst others, may make waiting in the open bus stop undesirable. Besides, it is arguably less of a pain point to be on board the bus, seated and enjoying the air-conditioner albeit slowly moving, compared to waiting at the bus stop with uncertainty on when a bus would actually arrive.

Whilst allowing defections out of goodwill and compassion for the needy is certainly worthwhile, without a mechanism for check and balance, this may be subjected to abuse since everybody would instinctively like to board immediately instead of wasting extra time waiting for the next bus. But if too many people defect, the system as a whole would fail to maintain its optimal configuration of buses with bus bunching being a repercussion --- defeating the original intention of the no-boarding policy. Such a situation is an example of the \emph{tragedy of the commons} \cite{Hardin68}.

\subsection{Inductive reasoning and bounded rationality}

To simulate people evaluating choices and making what they would individually perceive as their respective best action \cite{Bower81,Holland89}, we adopt the description of humans with inductive reasoning and bounded rationality presented by Ref.\ \cite{Arthur94} in studying the El Farol Problem. That description has subsequently been formalised into the \emph{minority game} \cite{Challet97,Challet98,Savit99,Cavagna99,Manuca00,Moro04}. Our mathematical representation of the decision-making people (or agents) is based on Ref.\ \cite{Challet97}, which we now describe.

When a bus announces that the no-boarding policy is activated, here in this paper, each person who would otherwise normally be allowed to board but denied in Ref.\ \cite{Vee2019b} would be given the options to cooperate (i.e. not board, and obediently wait for the next bus) or defect (i.e. defy the no-boarding rule and board anyway). The way they make their choices is determined as follows. We represent them as independent agents each endowed with $s$ different strategies and memory $m$. These $s$ strategies are ideally distinct for different agents, since different people would behave differently with their own beliefs and ideas. Nevertheless, coincidentally similar strategies are allowed. In fact, if the number of agents $N$ is way more than there are available strategies (which is determined by $m$, i.e. $2^m$ strategies, see Refs.\ \cite{Challet98,Savit99} for discussions on this), then some agents \emph{must} be sharing at least one strategy. The memory records the $m$ most recent past results of the winning choices, i.e. whether the cooperators are winners or the defectors are winners. Then, for each such $2^m$ combinations of $m$ binary historical outcomes, a strategy would specify the next action of whether to cooperate or defect. A strategy is thus a set of maps, one map for each of these $2^m$ possible combinations to an element of $\{$cooperate, defect$\}$ to be made in the next round. A different strategy would map each of those $2^m$ combinations into a possibly different element of $\{$cooperate, defect$\}$. The performance of each of the $s$ strategies is tracked based on their predictions and the actual outcomes, and a strategy gains a ``virtual point'' for correctly predicting the next outcome. These virtual points track how well each strategy is predicting the next outcome regardless whether they are used or not in making the actual choice. The current best performing strategy (i.e. with the highest virtual point) would be used to make the actual next decision. If multiple strategies are tied on virtual points, the actual one to use is decided randomly. Thus, this adaptivity property allows agents to learn which strategy amongst their $s$ possible ones is so-called ``current best'' from their individual perspectives.

\subsection{Winners are those in the minority group, collective cooperation, herding behaviour}

Note that in Refs.\ \cite{Arthur94,Challet97}, the winning group (cooperators or defectors) are determined by the \emph{minority group}, i.e. the side with fewer people. This rule presents itself with a natural feedback mechanism that never settles into any permanently desired group. If cooperators are the winners right now, then more defectors would like to be cooperators. But this would turn cooperators into the majority and defectors would become the winning group.

Such a minority feedback has led physicists to draw comparisons with physical systems possessing quenched disorder and phase transitions \cite{Savit99,Challet99,Manuca00,Marsili00,Challet00,Sherrington75,Moro04,Ghosh12,Chakra15}. For instance, the minority game has different phases, where in one such phase the best strategies for a significant number of agents, respectively, are \emph{frozen}, i.e. these are always their best respective strategies and these agents never switch into a different strategy. Ref.\ \cite{Challet99} argued that this is akin to symmetry breaking of a spin system when the control parameter exceeds a critical value, analogous to spontaneous magnetisation. Such correspondences with physical systems have fruitfully opened up the use of statistical mechanical techniques being applied to the minority game \cite{Marsili00,Challet00,Sherrington75}, with applications even to financial markets \cite{Zhang98,Challet00b,Challet01,Marsili01,Jeff01,Bou01,Challet01,Marsili02} and other problems on resource allocations \cite{Ghosh12,Chakra15}.

A key revelation from research in the minority game is that there is a regime where the entire community exhibits incredible \emph{collectively cooperative behaviour} even though each agent is only interested in self-gain. Apart from this surprising positive global behaviour, another astonishing result arises when the number of agents, $N$, is much greater than the available independent strategies, $2^m$: \emph{herding behaviour} emerges where many agents with similar strategies behave as a crowd who take the same action. This is highly undesirable as it would lead to a skewed outcome, with a small minority group containing few winners. In terms of resource utilisation, this means that there are many people who could have benefited but missed out because they joined the ``majority bandwagon''. Quantitatively, the variance from optimal utilisation per number of agents varies as a power law with respect to $2^m/N$ in the regime where $N\gg2^m$ [see Fig.\ \ref{fig2}(a) in Section 2].

In the bus problem that we are considering, this is also a resource allocation optimisation problem as the bus system strives to enhance the efficiency in serving commuters. The no-boarding policy essentially imposes a limit on the capacity, i.e. it is a bounded resource, which is being competed by the waiting commuters at the bus stop.

\subsection{Who are winners in cooperating or defecting the no-boarding bus?}

For the bus system, on the other hand, there is no clear, obvious, nor a unique generalised meaning for cooperators or defectors being in a ``minority group''. Does ``less defectors compared to cooperators'' imply that defectors ``win''? Why should a na\"{i}ve ``less defectors compared to cooperators'' allow defectors to declare victory? After all, a so-called successful implementation of the no-boarding policy aims to minimise defections, i.e. ``zero defections is ideal'', from the point of view of the bus system. Hence, a key part of this paper is to define and formalise what it means for the cooperators or defectors to win, instead of just counting the numbers in each group and seeing which has less people. One important aspect of the original minority game was to optimise usage of the resource, i.e. whilst the minority group wins, the system as a whole would be considered as ``optimal'' if the wastage or deviation from the ideal capacity is minimised. Once again offhand, it is not directly straightforward what this would be for the bus system, or if such a notion is even applicable here.

We note that Ref.\ \cite{Cavagna99} found that the actual historical outcomes of winners is not crucial in cultivating the emergence of collective cooperation. Instead, \emph{any} exogeneous piece of information is sufficient to generate that community-wide learning. Therefore, as long as the bus system systematically decides who the winners are and this information is made clear to all agents (for instance, defectors get away and thus ``win'' this time, or they are all punished with a fine for defecting and so cooperators are ``winners'' at another time), this would probably also be true for the no-boarding bus where people can choose whether to cooperate or defect. In other words, winners being decided by the minority rule can be replaced by other winning criteria, whilst maintaining the key feature of the community's collective cooperation. We will verify this in this paper.

\subsection{More total agents than those who are actually playing each round}

The classical minority game sets a fixed number of agents who repeatedly play amongst all of them. But in the real world, this is definitely not the case. Why should everybody always play each round? Some people may take a break or play only occassionally. In the original El Farol Problem \cite{Arthur94}, it is arguable that a realistic situation may be that there are say 200 people, but on average only about 100 of them \emph{would actually consider} whether or not to go to the bar and compete for the 60 available places, with the remaining people taking a break from playing.

In fact in the bus system, the actual number of people boarding a bus each time is not even fixed! The total number of people using the bus system is overwhelmingly more than the number of people boarding each time, with even fewer who are actually boarding when the no-boarding policy is activated. In view of this, the classical minority game needs to be extended to a situation where there are necessarily \emph{more agents in the overall pool of people than those who are actually playing each round}; as opposed to a fixed number of agents who always play against each other every round.

The minority game has been applied to study behaviour and phenomena associated with financial markets \cite{Zhang98,Challet00b,Challet01,Marsili01,Jeff01,Bou01,Challet01,Marsili02}. A financial market involves buying or selling commodity, where some agents would only react and execute an action when they feel that the time is right. This is thus a departure from the classical minority game where not every agent plays every round. Several models and modifications to the classical minority game have been proposed, for instance Ref.\ \cite{Jeff01} introduces a minimum threshold on an agent's predictability rate before going ahead with an actual action of buying or selling. This reflects real market agents who invest some time to observe the prices to gain confidence with their intuition. Alternatively, some agents (referred to as ``speculators'') only play when they receive enough information in a financial market, whilst others (referred to as ``producers'') always play \cite{Challet00b,Challet01}. Another model for agents' behaviours is that they trade at different time scales \cite{Marsili02}. These models have achieved success in explaining certain phenomena typically occurring in a financial market, like the emergence of stylised facts due to the presence of speculators \cite{Challet01}, as well as long-range volatility correlation effects \cite{Bou01}. In fact, Refs.\ \cite{Challet00b,Marsili02} found that the phase transition between the informationally efficient and inefficient phases still exists, and showed how the critical value for the control parameter varies with additional parameters in their models.

The herding behaviour of the agents in the classical minority game is a major cause of concern for the bus system. If in some successive rounds a massive crowd decides to defect, this may slow down the bus too much and nullify the no-boarding policy --- leading to bus bunching. But since the bus system is not quite like the classical minority game, we need to study it explicitly and find out what happens. Unlike a financial market, people boarding a bus do not actually decide by themselves if they ``want to play the no-boarding game''. A bus would activate no-boarding when it is too slow and bunching is imminent, beyond the control of an individual commuter. Therefore, in contrast to those models describing agents in a financial market, we consider a pool of agents where a \emph{random} subset of them play the no-boarding game, whenever a bus activates it in a bus loop service.

In the next section, we recapitulate on the classical setup and features of the minority game, stating the key results as well as the behaviour of the agents in adapting to the optimal situation and how efficiency of resource usage depends on factors like number of agents and memory \cite{Challet97,Challet98,Savit99,Manuca00,Moro04}. We also consider an ``open'' minority game, where there are $2N$ total agents with only $N$ of them (randomly selected) who are actually playing each round. Then in Section 3, we investigate several situations for the bus system where a bus allows defectors to go through ``victoriously'' when there are few of them, but is capable of correcting the situation and ``punishes'' defectors when the defection level becomes high. Interestingly, the agents with such inductive reasoning and bounded rationality are indeed capable of adapting to the rules and optimising according to different situations at a global scale.


\section{Classical minority game}

\begin{figure}
\centering
\includegraphics[width=13cm]{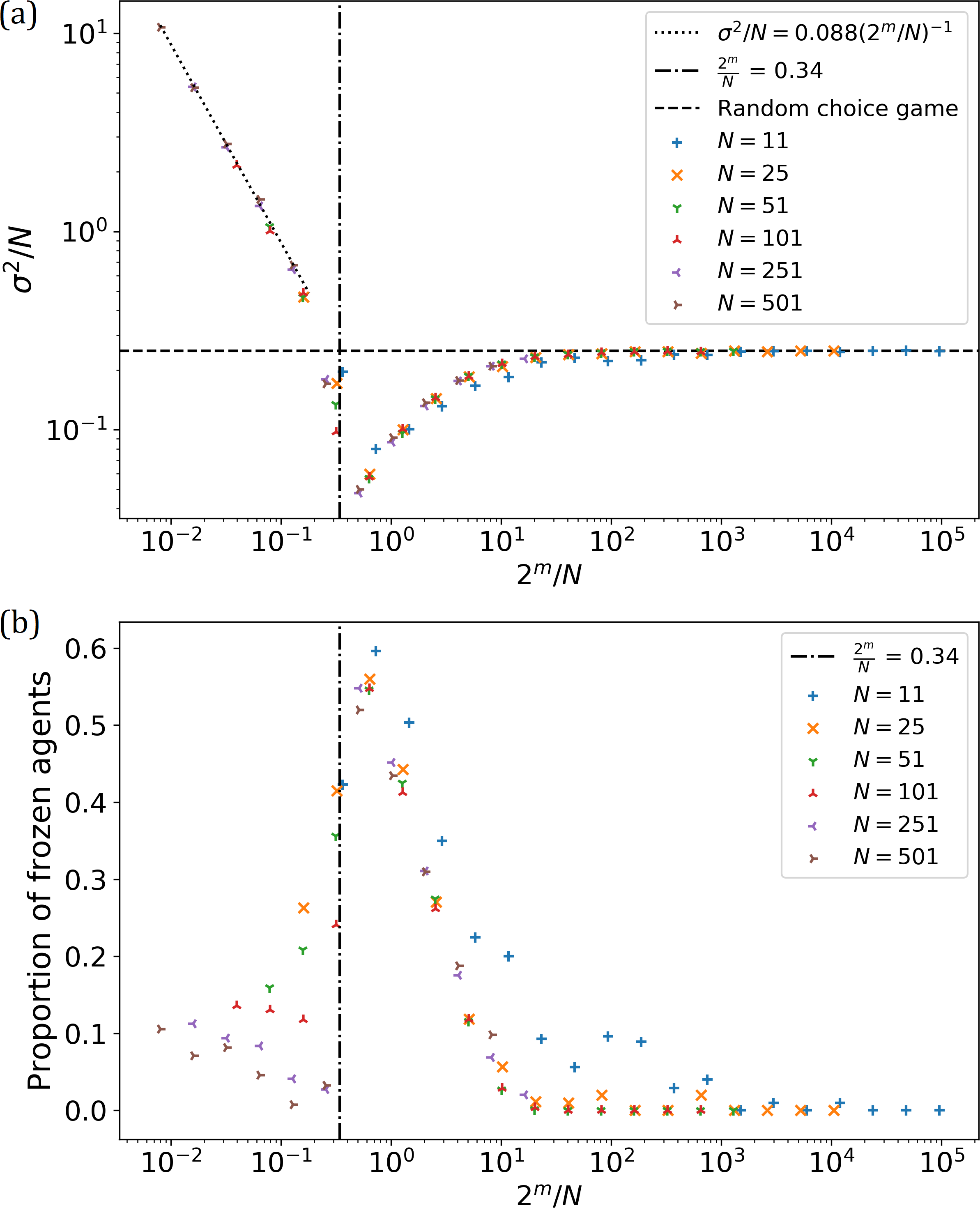}
\caption{(a) The universal curve in the classical minority game. (b) The proportion of agents who stick with one of their two strategies.}
\label{fig2}
\end{figure}

Fig.\ \ref{fig2}(a) shows a classical result of the minority game, viz. a graph of the variance from the ideal scenario [where the minority group size is $(N-1)/2$] per number of agents, $\sigma^2/N$, versus $2^m/N$, where $m$ is the memory and $N$ is the number of agents \cite{Challet97,Challet98,Savit99,Manuca00,Moro04}. We produce this graph by creating $N$ agents, each endowed with $s=2$ strategies which map the past $m$ most recent winning outcomes to a next action to be taken, as described in the Introduction. Each of these agents plays for $20000$ rounds, and measurements are taken in the last $30\%$ of the latter rounds to ensure that the transient part of the simulation has subsided. This entire simulation is repeated $100$ times where the agents are randomly initialised with different $s=2$ strategies, and the average over these 100 runs is computed.

\emph{Note}: In all our simulations in this paper, we additionally carry out extensive simulations at half the number of rounds that are set, respectively, for each run. For example, if our reported results use $20000$ rounds, then we also carry out separate runs with $10000$ rounds. In all our reported results, the corresponding runs at half the number of rounds display essentially identical results as those for the stated number of runs. This implies that the stationary state has been attained long before half the number of runs.

A game comprises $N$ agents, typically an odd number, where each agent selects one of two options based on their current best performing strategy. The minority rule determines the winning group (which is well-defined, since $N$ is odd). The so-called best configuration for the whole community is when there are $(N-1)/2$ agents selecting one choice, with $(N+1)/2$ agents selecting the other. The former is the winner, but the point here is that \emph{the number of winners is maximised} --- which is the notion of optimal use of the resource since the capacity of winners is $(N-1)/2$ in each round. These same $N$ agents play the game repeatedly, until the system settles into a state where the mean number of agents choosing one of the choices is about $N/2$, with variance $\sigma^2$ that depends on $m$ and $N$, according to Fig.\ \ref{fig2}(a). In fact, Fig.\ \ref{fig2}(a) is a \emph{universal curve} which holds true for any $m$ and $N$.

We summarise the known results and features of the classical minority game \cite{Challet97,Challet98,Savit99,Manuca00,Moro04}:
\begin{enumerate}
\item There are two phases separated by $2^m/N\sim0.34$.

\item For the phase $2^m/N\lesssim0.34$, there are relatively fewer available strategies compared to the number of agents. Therefore, it is inevitable that some or many agents share similar strategies. Consequently, clusters of agents with similar strategies behave like herds or crowds, and the community as a whole performs badly. The deviation from optimal resource usage is high. The increase in $\sigma^2/N$ as $2^m/N$ decreases is a power law.

\item On the other hand, for $2^m/N\gtrsim0.34$, there are relatively more strategies available compared to the number of agents. Thus, herding behaviour is avoided and the community as a whole behaves much better than the random choice game (i.e. where each agent makes a choice with probability $0.5$).

\item As the memory $m$ gets larger, each agent tracks more information and becomes more complicated. This increased complexity turns out to make them behave more randomly. Thus, the universal curve approaches the random choice game asymptotically as $m\rightarrow\infty$, but the community is generally still performing better than the random choice game.

\item As shown in Fig.\ \ref{fig2}(b), the proportion of agents who stick to one of their two strategies shows a peak in the region where the entire community behaves most cooperatively. That proportion is substantial, with about half of the total number of agents experiencing a frozen strategy.

\item Fig.\ \ref{fig2} shows the situation where each agent is endowed with $s=2$ strategies. With larger $s$, the added complexity tends to diminish the region where the curve is below the random choice game in Fig.\ \ref{fig2}(a), and the universal curve approaches the random choice game as $s$ increases.
\end{enumerate}

\subsection{Open minority game}

\begin{figure}
\centering
\includegraphics[width=13cm]{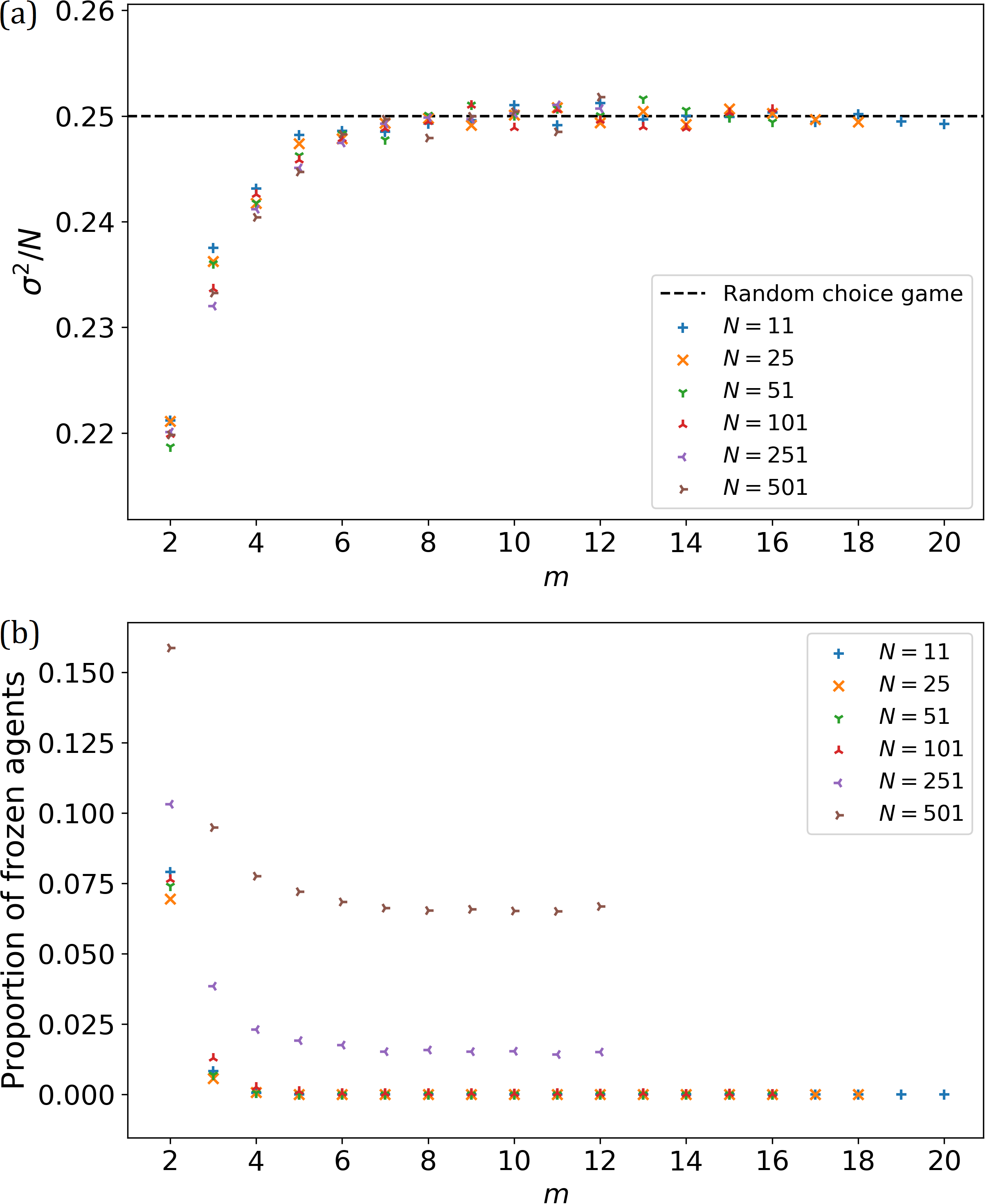}
\caption{(a) The universal curve in the open minority game. (b) The proportion of agents who stick with one of their two strategies.}
\label{fig3}
\end{figure}

Let us now consider the situation where there is a pool of $2N$ agents. Each round, $N$ agents randomly selected from that pool of $2N$ agents would participate in the minority game whilst the remaining $N$ stay out for that round. The rest of the details are identical to those in the classical minority game presented above. Fig.\ \ref{fig3} displays our simulation results. Note the important differences between this \emph{open} minority game as compared to the classical minority game where in the latter all agents always participate in the game during each round:

\begin{enumerate}
\item The plots in Fig.\ \ref{fig2} for the classical minority game are in log-log and semi-log scales, respectively, with the $x$-axis being with respect to $2^m/N$. The plots in Fig.\ \ref{fig3} for the open minority game on the other hand are in the usual linear scale, with the $x$-axis being with respect to $m$ (independent of $N$).

\item In Fig.\ \ref{fig3}(a), the variance per number of actual competing agents $\sigma^2/N$ decreases with decreasing $m$, but not as significantly as that in the classical minority game. Most strikingly, there is \emph{no herding behaviour} for regions of low $m$ in the open minority game, i.e. no power law increase in $\sigma^2/N$ as $m$ decreases beyond the critical value of $2^m/N\sim0.34$ and no phase transition.

\item A universal curve in the open minority game appears to be between $\sigma^2/N$ and $m$, instead of $\sigma^2/N$ and $2^m/N$ for the classical minority game.

\item From Fig.\ \ref{fig3}(b), the proportion of frozen agents who stick to one of their two available strategies in the open minority game is significantly lower than that in the classical minority game, with no peak of the order of $\sim0.5$. Nevertheless, it appears to slightly increase with decreasing $m$ where the efficiency in terms of resource usage is better (i.e. where $\sigma^2/N$ is smaller).

\end{enumerate}

The absence of the herding behaviour when $N\gg2^m$ should not be a surprise in the open minority game in spite of the number of agents playing in each round, $N$, being much more than the available strategies, $2^m$. Each agent is not playing against the identical group of agents during each round but instead faces different agents whenever they play. Hence, there is no reason to expect the collective crowd or herding behaviour that exists in the classical minority game where they \emph{always} face the same group of agents. Whilst agents ``get used to each other'' in the classical version, they persistently face random opponents in the open version which continually alters their best strategies.

The open minority game is a more realistic description of real-world systems like people boarding a bus, than the classical version. For example, different groups of people would board the bus each time, and each person faces a different group of people each time.

\section{No-boarding buses with inductive reasoning and bounded rationality}

Let us now move on to deal with the problem of interest, viz. a bus system with a no-boarding policy. Our simulation environment for a bus loop system is based on that developed in Ref.\ \cite{Vee2019b}, with parameters tuned from a real university campus bus loop service \cite{Vee2019,Vee2019b}. A simplified setup with realistic parameters is to consider two buses serving one bus stop in a loop. We let both these buses move with a natural period of $12$ minutes (excluding time stopped at the bus stop), and people would arrive at the bus stop at a rate of $1$ person every $16$ seconds. At the bus stop, a bus would first allow people to alight, before boarding new passengers. The people load/unload at a rate of $1$ person per second. Consequently, each bus would pick up an average of $24$ people each time. The parameters for this setup is summarised in Table\ \ref{table1}. A different setup with $3$ buses is presented in the following subsection.

\begin{table}
\centering
\begin{tabular}{|l|c|c|}
\hline
Quantity & First setup & Second setup\\
\hline
Number of buses & $2$ & $3$\\
\hline
Number of bus stops & $1$ & $1$ \\
\hline
Total number of agents & $50$ & $100$ \\
\hline
Natural period of each bus (minutes) & $12$ & $12$\\
\hline
Interarrival time of people at a bus stop (seconds) & $16$ & $10$\\
\hline
Number of people boarding/alighting bus (per second) & $1$ & $1$\\
\hline
No-boarding angle, $\theta_0$ & $120^\circ$ from behind & $150^\circ$ from ahead\\
\hline
Length of one run (number of revolutions) & $10000$ & $10000$\\
\hline
Number of runs to obtain average & $100$ & $100$\\
\hline\end{tabular}
\caption{The parameters used in our simulations.}\label{table1}
\end{table}

As it is, the two buses would quickly bunch into one single unit. In this case, the average waiting time of people at the bus stop for a bus to arrive is about $6$ minutes and $11$ seconds. To prevent this, a ``slow'' bus would implement the no-boarding policy, i.e. it only allows alighting and then leave, if its phase difference as measured from the bus behind it (or the bus immediately behind it if there are more than two buses) becomes less than some critical $\theta_0$. (See the following subsection for an alternative criterion for implementing the no-boarding policy, based on measuring the phase difference from the bus immediately \emph{ahead}.) Note that since the buses go in a loop, one can map this loop isometrically to the unit circle where the notion of a phase on the circle ($0^\circ$ to $360^\circ$) is well-defined, and we can speak of the phase difference between two buses on this unit circle. We set $\theta_0=120^\circ$ so that if the phase difference between two buses gets smaller than that, the leading bus (which is ``too slow'') would disallow boarding at the bus stop, leaving the people there to be picked up by the trailing bus (to slow it down, since it is ``too fast'').

Ref.\ \cite{Vee2019b} has established significant improvements due to the no-boarding policy in preventing bus bunching and dramatically reducing the waiting time of people at the bus stop for a bus to arrive. With this setup where $\theta_0=120^\circ$, the average waiting time is only about $3$ minutes and $47$ seconds, an improvement of almost $40\%$ from the situation without the no-boarding policy. Instead of mandatorily enforcing the no-boarding policy when the phase difference drops below $\theta_0$, here we allow the people waiting to board to choose whether to cooperate or defect the no-boarding policy, when it is activated. Each person who would normally be allowed to board would decide for themselves, based on the inductive reasoning and bounded rationality described above. Then those who decide to defect would proceed to board as usual, whilst those who decide to cooperate would remain at the bus stop and wait for the next bus.

In this setup, the actual number of people who have to ``play the no-boarding game'' (i.e. who are faced with no-boarding, but given the choices to cooperate or defect) varies each time, from as low as $1$ person up to occasionally nearly $30$ people. The mean number is around $10$ to $16$ people (Table\ \ref{table2}, below), depending on the actual learning and behaviour of the agents (e.g. their memory $m$). These people who face the no-boarding policy are the excess from the average of $24$ people to be picked up by each bus, who are accumulated at the bus stop due to the bus being ``slow''. We create a pool of $50$ agents, and each round only a subset of these $50$ agents ``play the no-boarding game''. This mimics the real situation where different subsets of the population board the bus at different times, with an even smaller subset who actually has to face the no-boarding policy. Nevertheless, we keep the overall pool to the relatively small number of $50$ agents, to avoid excessively lengthy training runs of the simulations required to weed out the transient part in order to allow the agents to learn their best respective strategies.

For a criterion on determining the winning group, suppose the bus system allows a fixed number of defectors to board when the no-boarding policy is implemented during that stop. This number is arbitrarily decided by the bus system: Perhaps it could set a larger limiting number during lull times when the pressure on bus bunching is weaker and a smaller limiting number during busy times when a bus would have to stop longer to serve more passengers. If the number of defectors is within this limit, then they get away without any punishment. In this sense, the defectors are deemed as winners, whilst the cooperators are losers since they apparently ``wasted their time for nothing'' when they obeyed and waited for the next bus. On the other hand, if the number of defectors exceeds the prescribed limit, then all those who defected that round are punished and charged a (possibly hefty) fee whilst the cooperators of that round are given a rebate for obediently following the rule. This information on how many defectors are allowed each time, however, is not announced to the agents as it is only meant for the bus system to decide on the winning group. Therefore, each agent has to individually weigh the pros and cons of defecting the no-boarding rule, if it is activated:

\vskip 0.5cm

\emph{Cooperating is incentivised by possibly being rewarded with a rebate in exchange for the extra waiting time for the next bus, if there are too many defectors. But a cooperator may waste their time for nothing, if there are too few defectors.}

\vskip 0.5cm

\emph{Defecting is incentivised by saving time and possibly getting away with it, if there are not too many defectors. But a defector risks incurring a fine, if there are too many defectors.}

\vskip 0.5cm

\begin{figure}
\centering
\includegraphics[width=13cm]{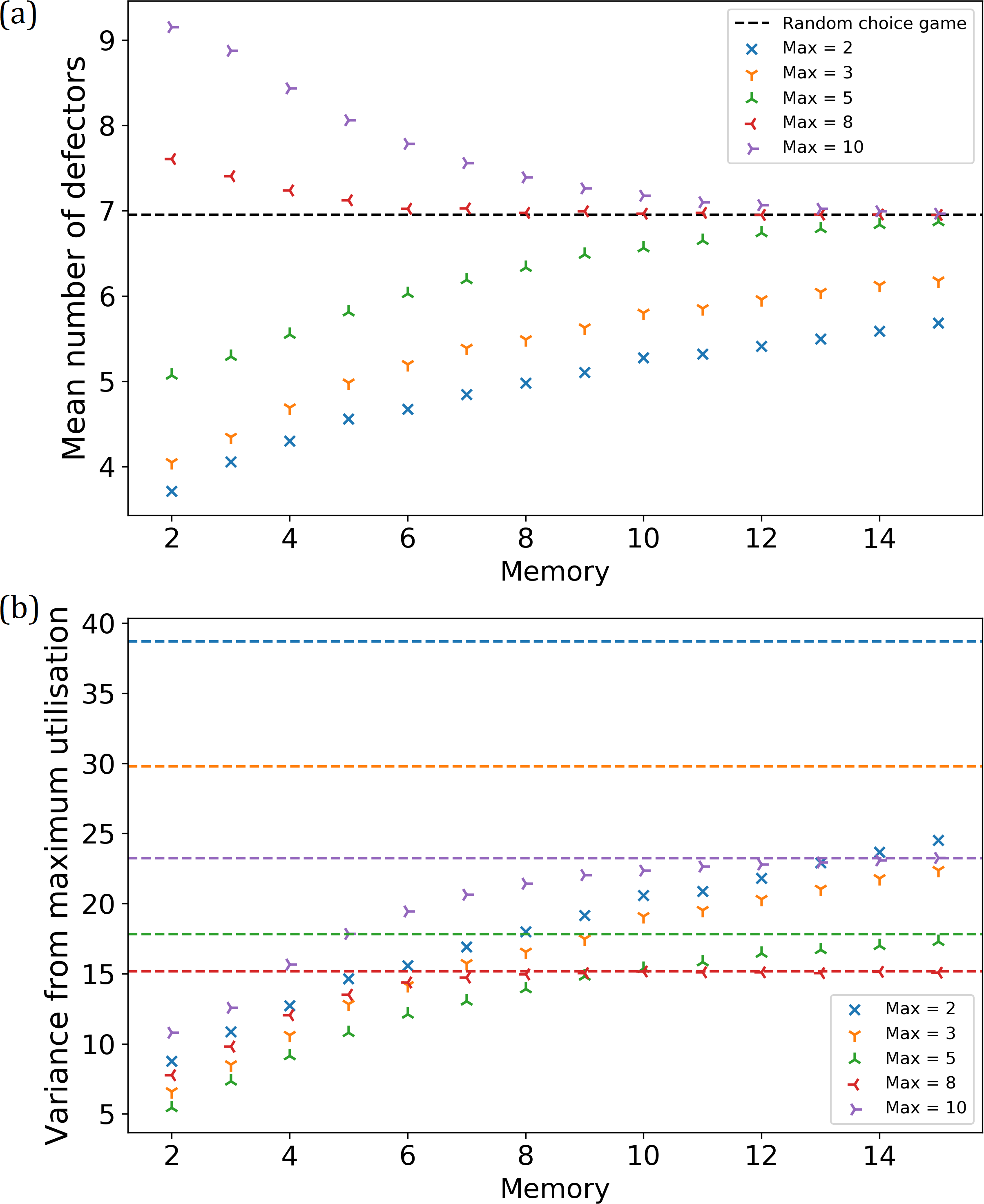}
\caption{Various fixed limit at 2, 3, 5, 8, 10 allowed defectors, respectively: (a) Mean number of defectors versus memory. (b) The variance from maximum utilisation of the defection capacity versus memory.}
\label{fig4}
\end{figure}

We shall set various fixed limits: a maximum of $2$, $3$, $5$, $8$, or $10$ defectors, and test this in our simulations one at a time, with the bus system running for $10000$ revolutions around the loop. Furthermore, we also run control simulations where instead of modelling agents with inductive reasoning and bounded rationality, all agents behave randomly --- i.e. they flip a fair coin to decide whether to cooperate or defect the no-boarding rule. Each case is repeated $100$ times to obtain the average. Fig. \ref{fig4} summarises our simulation results.

\begin{figure}
\centering
\includegraphics[width=13cm]{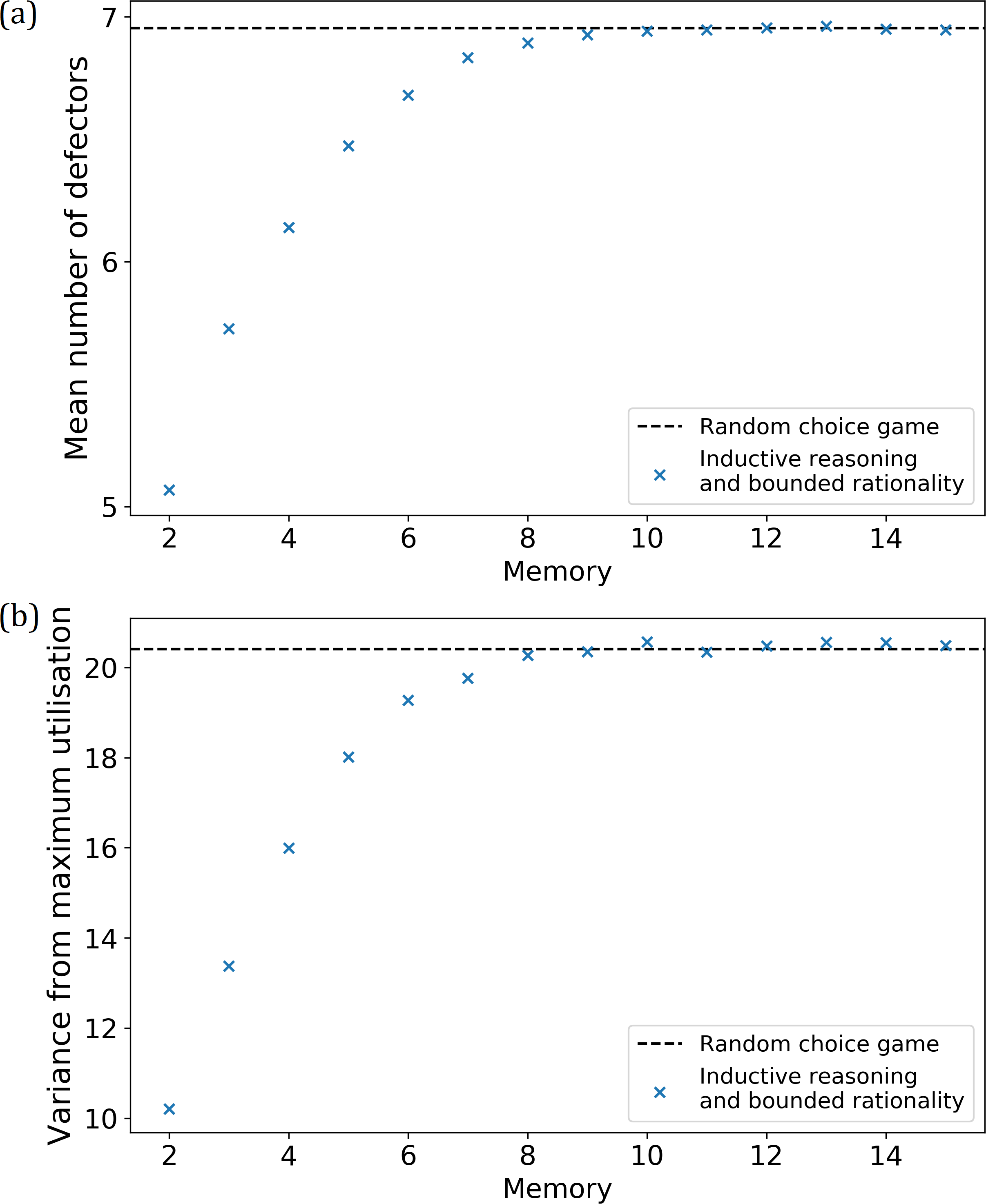}
\caption{The limiting number of defectors is set as the actual number of defectors in the previous round: (a) Mean number of defectors versus memory. (b) The variance from maximum utilisation of the defection capacity versus memory.}
\label{fig5}
\end{figure}

Additionally, we also let the limiting number of defectors be variable: The maximum number of defectors in the next time the no-boarding policy is activated is taken as \emph{the actual number of defectors in the previous time} the no-boarding policy is activated. This removes the arbitrary fixing of the limiting number of defectors, replaced by the actual number of defectors in the last round. Fig.\ \ref{fig5} is the corresponding plots to Fig.\ \ref{fig4} in this variable case.

In each case where defection is allowed, the bus system reasonably maintains its performance where the two buses do not bunch. The allowance of defection only slightly increases the global average waiting time of people at the bus stop for a bus to arrive. Our simulation results show that agents behaving with inductive reasoning and bounded rationality are capable of adapting to the winning criteria, regardless of what those criteria are. In each of the maximum number of defectors set by the bus system, the agents are able to co-evolve their best strategies such that the mean number of defectors each round the no-boarding policy is activated approaches the maximum number being set by the bus system, \emph{in spite of the fact that the agents themselves do not actually know what the limit is!} The only information that each agent receives and remembers is the last $m$ outcomes that he experienced, which he is able to track by himself since he knows whether he won or lost (viz. getting a rebate or not when cooperating; paying a fine or not when defecting). Yet the entire community optimises the variable resource usage, with lower memory being more effective (lower variance from maximum utilisation) since larger $m$ increases the agents' complexity in their strategies and they approach the random choice game. The simulation results also clearly show the character of the open minority game, with no herding behaviour since each agent continually faces different groups of agents which alter their best strategies.

\begin{table}
\centering
\begin{tabular}{|c|c|c|c|c|}
\hline
Allowed defectors & No. of defectors & No. of agents & \% defectors & Increase in mean waiting time\\
\hline
$2$ & $3.7$ & $10.9$ & $34.2\%$ & $1.9\%$ or $4.3$ seconds \\
$3$ & $4.1$ & $10.9$ & $37.1\%$ & $2.2\%$ or $5.0$ seconds \\
$5$ & $5.1$ & $11.7$ & $43.2\%$ & $2.2\%$ or $5.0$ seconds \\
$8$ & $7.6$ & $14.4$ & $52.8\%$ & $2.9\%$ or $6.5$ seconds \\
$10$ & $9.1$ & $16.1$ & $56.9\%$ & $3.5\%$ or $7.9$ seconds \\
Previous number & $5.1$ & $11.9$ & $42.7\%$ & $2.2\%$ or $5.0$ seconds \\
\hline
Random choice game & $7.0$ & $14.0$ & $50.0\%$ & $2.5\%$ or $5.8$ seconds\\
\hline
\end{tabular}
\caption{The case where the memory of the agents is $m=2$, for the various numbers of allowed defectors set by the bus system: The mean number of defectors, mean number of agents given the choices to cooperate or defect the no-boarding policy, mean proportion of defectors, and the increase in mean waiting time as compared to the case where the no-boarding policy is mandatory. The last line is where each agent randomly decides to cooperate or defect.}\label{table2}
\end{table}

Note also that by adapting to the limiting number of defectors set by the bus system, the entire community is capable of co-evolving their proportion of defection rates. For example with $m=2$ (Table\ \ref{table2}), when only $2$ defectors are allowed, there are $10.9$ people, on average, given the choices to cooperate or defect the no-boarding policy, with about $3.7$ defectors on average, i.e. a defection rate of $34.2\%$. On the other hand when the bus system allows $10$ defectors, there are $16.1$ people, on average, given the choices to cooperate or defect the no-boarding policy, with $9.1$ defectors on average, i.e. a defection rate of $56.9\%$. Allowing more defectors would slightly increase the global average waiting time, which slightly slows down the ``slow'' bus and thus slightly increases the excess people who would face the no-boarding policy. This is why allowing more defectors would increase the average number of agents who ``play each round''. Table\ \ref{table2} shows that the agents are able to dynamically adjust the community's defection rate to the various winning criteria, not just the $50-50$ rate of defections that would arise from the minority rule for deciding the winning group. The adaptability weakens with increasing memory $m$, since the agents' complexity increases and would behave more randomly, just like the classical and open minority games where the variance from ideal utilisation grows towards that of the random choice game. The increases in mean waiting time are less than $10$ seconds in all cases, compared to $3$ minutes and $47$ seconds where no-boarding is mandatorily implemented with no option to defect. This is only a small cost, but gives people with urgency to board the option to do so.

\subsection{3 buses}

\begin{table}
\centering
\begin{tabular}{|c|c|c|c|c|}
\hline
Allowed defectors & No. of defectors & No. of agents & \% defectors & Increase in mean waiting time\\
\hline
$2$ & $4.8$ & $15.3$ & $31.4\%$ & $1.4\%$ or $2.2$ seconds \\
$5$ & $5.4$ & $15.7$ & $34.4\%$ & $1.9\%$ or $2.9$ seconds \\
$8$ & $6.2$ & $15.9$ & $39.0\%$ & $2.3\%$ or $3.6$ seconds \\
$15$ & $10.8$ & $20.8$ & $51.9\%$ & $7.0\%$ or $10.8$ seconds \\
$18$ & $17.1$ & $31.8$ & $53.8\%$ & Bus bunching \\
Previous number & $5.6$ & $15.7$ & $35.7\%$ & $2.3\%$ or $3.6$ seconds \\
\hline
Random choice game & $9.6$ & $19.6$ & $49.0\%$ & $5.6\%$ or $8.6$ seconds\\
\hline
\end{tabular}
\caption{The corresponding table for Table\ \ref{table2}, for the second setup with $N=3$ buses.}\label{table3}
\end{table}

\begin{figure}
\centering
\includegraphics[width=13cm]{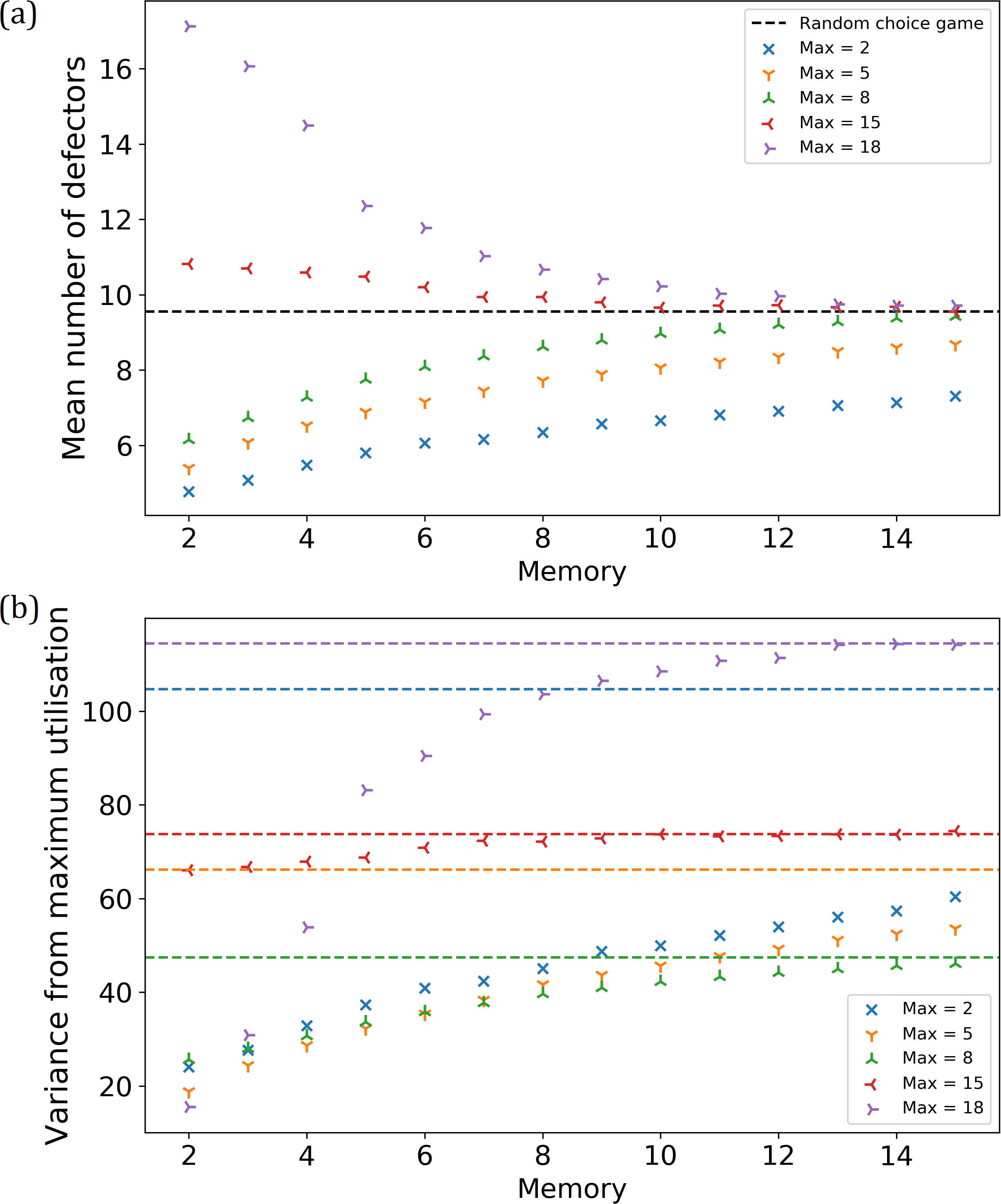}
\caption{Various fixed limit at 2, 5, 8, 15, 18 allowed defectors, respectively: (a) Mean number of defectors versus memory. (b) The variance from maximum utilisation of the defection capacity versus memory.}
\label{fig6}
\end{figure}

\begin{figure}
\centering
\includegraphics[width=13cm]{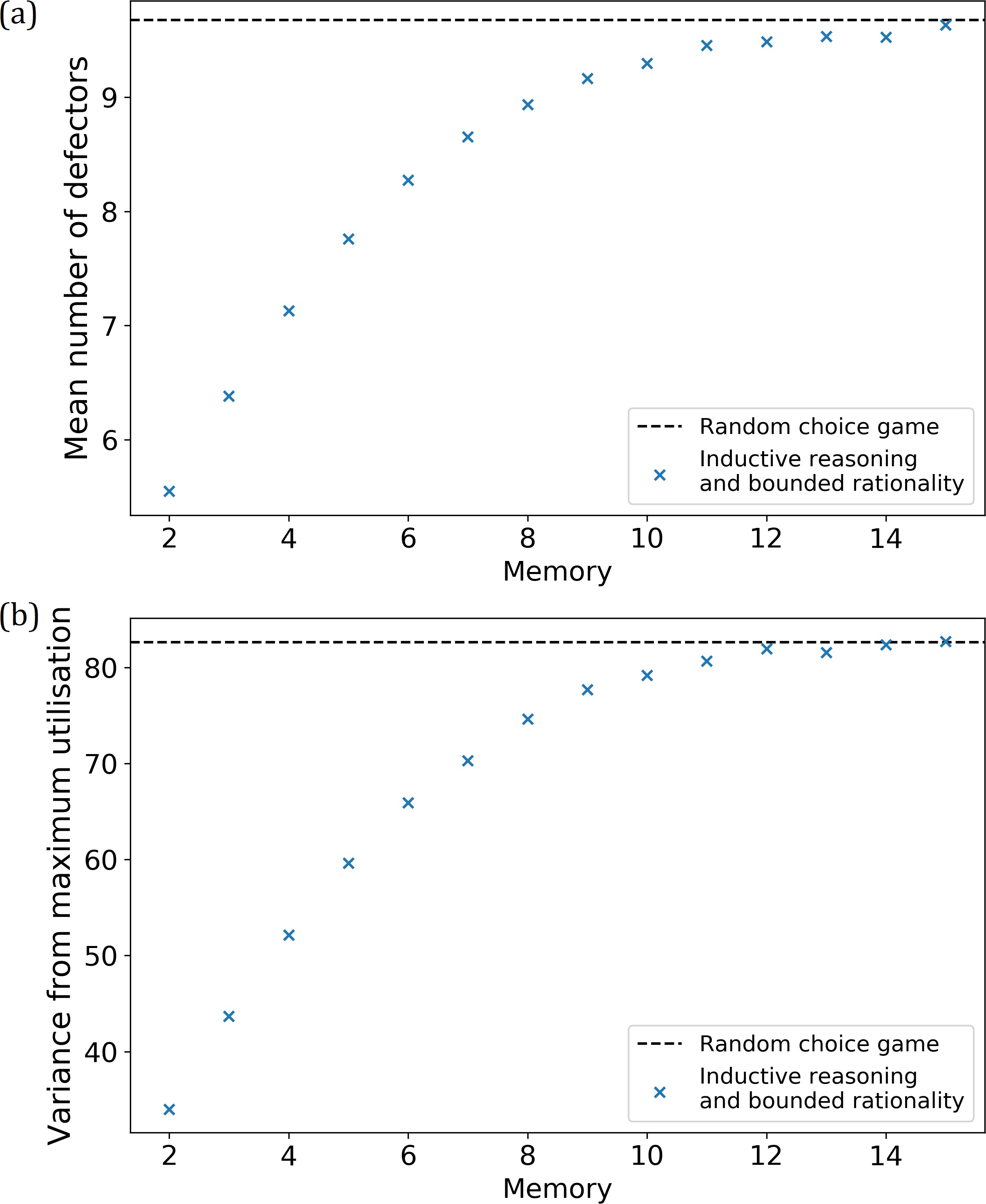}
\caption{The limiting number of defectors is set as the actual number of defectors in the previous round: (a) Mean number of defectors versus memory. (b) The variance from maximum utilisation of the defection capacity versus memory.}
\label{fig7}
\end{figure}

We repeat the entire set of simulations on a setup with 3 buses serving 1 bus stop in a loop, with a higher rate of 1 person arriving at the bus stop every 10 seconds. Without the no-boarding policy, all three buses would bunch into one single unit and the average waiting time at the bus stop is 6 minutes and 12 seconds. Here, a bus implements the no-boarding policy if its phase difference measured from the bus \emph{ahead} of it is \emph{greater} than $\theta_0=150^\circ$, which implies that it is ``too slow''. (Note that in Ref.\ \cite{Vee2019b}, there are two different criteria to implement the no-boarding policy by measuring the phase difference from the bus immediately ahead or immediately behind.) Consequently, the three buses remain reasonably staggered and the average waiting time is 2 minutes and 34 seconds --- almost $60\%$ improvement. Each bus carries an average of 26 people. The setup parameters are stated in Table\ \ref{table1}, and the corresponding results are summarised in Table\ \ref{table3} as well as Figs.\ \ref{fig6} and \ref{fig7}. These results are qualitatively similar to those presented and discussed for the setup with 2 buses. The case with a limit of $18$ defectors is an extreme situation where too many defectors are allowed, leading to bus bunching.

\section{Concluding remarks}

The application of the inductive reasoning and bounded rationality description of agents to the no-boarding buses suggests that with the no-boarding policy improving the efficacy of the bus system \cite{Vee2019b}, allowing a fixed number of defectors each time the policy is activated is sustainable since the agents are able to adapt to making use of the available resource, without too much drawback on the bus system's performance. It is also a fitting example of an ``open'' system where only a subset of the overall pool of agents actually play the ``game'' in each round, hence there is no emergence of herding behaviour even though many agents may possibly be sharing similar strategies. This is because in such an ``open'' system, each agent faces different groups of opponents each time and therefore continually co-evolve their best strategies. Furthermore, we see similar features to the open minority game presented in this paper, even though the winning criterion for the bus system is not determined by the group with a smaller number of people.

Crucially, the absence of the herding behaviour assures the feasibility of the no-boarding policy with allowance for defections. The variance from the prescribed limiting number of defectors for a group of agents with inductive reasoning and bounded rationality being smaller than that for a group of randomly behaving agents (Figs.\ \ref{fig4} to \ref{fig7}), with no blowing up [unlike in the classical minority game when $N\gg2^m$, Fig.\ \ref{fig2}(a)], implies that the number of defectors reasonably hovers near the prescribed limit and it is unlikely that there would arise formation of huge crowds of people who decide to defect. Hence, the bus system is protected from the situation where there are surges of people defecting which may slow down the ``slow'' bus too much to an extent that bus bunching occurs, nullifying the intention of the no-boarding policy. This however, may not be the case if it is the same group of agents who always play repeatedly with each other as in the classical minority game, which does possess the herding regime. It is also unlike some models for financial market which still possess a phase transition \cite{Challet00b,Challet01} since commuters face the no-boarding policy by chance instead of getting to choose when to ``play the game''.

To our knowledge, there has not been any study on a no-boarding policy implemented by a bus system, other than a recent work in Ref. \cite{Vee2019b}. Perhaps a mandatory no-boarding policy may appear as being strict against people who need service urgently, and thereby dismissed as unfeasible. However, with Ref. \cite{Vee2019b} showing significant improvement in the bus system's efficiency, together with this paper's result that a no-boarding policy with options for the people to cooperate or defect giving rise to a only negligible effect on the bus system, perhaps such a policy may indeed be viable for a real-world implementation. In the event of a rare excessive number of defectors, the bus may then enforce a mandatory no-boarding and leave for that round to prevent imminent bus bunching. Also a future extension on this study may consider a punishment for defectors being proportional to the rate of defectors, i.e. the event of more numerous defectors would incur them a heavier fine, such that they are more strongly discouraged from defecting next time.

\begin{acknowledgments}
This work was supported by the Joint WASP/NTU Programme (Project No. M4082189) and the DSAIR@NTU Grant (Project No. M4082418).
\end{acknowledgments}

\section*{Data availability statement}

Data sharing is not applicable to this article as no new data were created or analysed in this study.

\bibliography{Citation}

\end{document}